\documentclass[11pt]{article}
	
	\newcommand{\blind}{0}
	
    \makeatletter
    \renewcommand\section{\@startsection {section}{1}{\z@}%
                                       {-3.5ex \@plus -1ex \@minus -.2ex}%
                                       {2.3ex \@plus.2ex}%
                                       {\normalfont\fontfamily{phv}\fontsize{16}{19}\bfseries}}
    \renewcommand\subsection{\@startsection{subsection}{2}{\z@}%
                                         {-3.25ex\@plus -1ex \@minus -.2ex}%
                                         {1.5ex \@plus .2ex}%
                                         {\normalfont\fontfamily{phv}\fontsize{14}{17}\bfseries}}
    \renewcommand\subsubsection{\@startsection{subsubsection}{3}{\z@}%
                                        {-3.25ex\@plus -1ex \@minus -.2ex}%
                                         {1.5ex \@plus .2ex}%
                                         {\normalfont\normalsize\fontfamily{phv}\fontsize{14}{17}\selectfont}}
    \makeatother
	
	\usepackage{amsmath}
	\usepackage{graphicx}
	\usepackage{enumerate}
	\usepackage{booktabs} 
	\usepackage{url} 
	\usepackage{subcaption}
	\usepackage{lipsum}
	\usepackage{multirow}
	\usepackage{xcolor}
	\usepackage{lineno}
	\usepackage{makecell}
	\usepackage{algorithm}
	\usepackage{algorithmic}
	\usepackage{booktabs}
	\usepackage{wasysym}
    \usepackage{amssymb}

\begin{document}
		
		\def\spacingset#1{\renewcommand{\baselinestretch}%
			{#1}\small\normalsize} \spacingset{1}

		\if0\blind
		{
			\title{\bf Generalized Radiograph Representation Learning via Cross-supervision between Images and Free-text Radiology Reports}
			\author{Hong-Yu Zhou$^{1,\dag}$, Xiaoyu Chen$^{2,\dag}$, Yinghao Zhang$^{2, \dag}$, Ruibang Luo$^1$,\\ Liansheng Wang$^{2,*}$, Yizhou Yu$^{1,*}$ \vspace{3mm}\\
			$^1$ Department of Computer Science, The University of Hong Kong,\\ Pokfulam, Hong Kong \vspace{1.5mm}\\
            $^2$ Department of Computer Science, Xiamen University,\\
              	Xiamen, China \vspace{1.5mm}\\
            $^{\dag}$ These authors contributed equally \vspace{1.5mm} \\
            $^*$ Corresponding authors: L.Wang (lswang@xmu.edu.cn) and \\Y.Yu (yizhouy@acm.org)}
			\date{}
			\maketitle
		} \fi

		

			
		
	\begin{abstract}
    Pre-training lays the foundation for recent successes in radiograph analysis supported by deep learning. It learns transferable image representations by conducting large-scale fully-supervised or self-supervised learning on a source domain. However, supervised pre-training requires a complex and labor intensive two-stage human-assisted annotation process while self-supervised learning cannot compete with the supervised paradigm. To tackle these issues, we propose a cross-supervised methodology named REviewing FreE-text Reports for Supervision (REFERS), which acquires free supervision signals from original radiology reports accompanying the radiographs. The proposed approach employs a vision transformer and is designed to learn joint representations from multiple views within every patient study. REFERS outperforms its transfer learning and self-supervised learning counterparts on 4 well-known X-ray datasets under extremely limited supervision. Moreover, REFERS even surpasses methods based on a source domain of radiographs with human-assisted structured labels. Thus REFERS has the potential to replace canonical pre-training methodologies.
	\end{abstract}
			
	\noindent%
	\spacingset{1.5} 

\section{Introduction}
Medical image analysis has achieved tremendous progress in recent years, thanks to the development of deep convolutional neural networks (DCNNs)~\cite{krizhevsky2012imagenet,simonyan2014very,szegedy2015going,he2016deep,huang2017densely}. At the core of DCNNs is visual representation learning~\cite{bengio2013representation}, where pre-training has been widely adopted and become the most dominant approach to obtain transferable representations. Typically, a large-scale dataset, also called the source domain, is first used for model pre-training. Transferable representations from the pre-trained model are further fine-tuned on other smaller downstream datasets, called target domains.

As one of the most general forms of medical images, radiographs have a great potential to be used in widespread applications~\cite{phillips2020chexphoto,taylor2018automated,carlile2020deployment}. In order to achieve (or at least approximate) radiologist-level diagnosis performance in these applications, it is common to transfer learned representations from natural images to radiographs~\cite{yosinski2014transferable,wang2017chestx}, and ImageNet~\cite{deng2009imagenet} based pre-training is most widely adopted in this context. On the other hand, self-supervised learning~\cite{chen2019self,zhou2021models,haghighi2021transferable,zhou2020comparing} has attracted much attention in the community because it is capable of learning transferable radiograph representations without any human annotations. Both methodologies have been proven to be effective in solving medical image analysis tasks, especially when the amount of labeled data in the target domain is quite limited. However, in the first approach, there is an inevitable problem, which is the existence of domain shifts between medical and natural images. For instance, it is possible to introduce harmful noises from natural images as radiographs have a different pixel intensity distribution. As for self-supervised learning, to the best of our knowledge, there still exist clear performance gaps between radiograph representations learned through self-supervised and label-supervised pre-training. To avoid these problems, building large-scale annotated radiograph datasets for label-supervised pre-training becomes an essential and urgent issue in radiograph analysis.

Recently, radiologists and computer scientists have managed to build medical datasets for label-supervised pre-training at the size of hundreds of thousands of images, such as ChestX-ray~\cite{wang2017chestx}, MIMIC~\cite{johnson2019mimic2} and CheXpert~\cite{irvin2019chexpert}. To acquire accurate labels for radiographs, these datasets often rely on a two-stage human intervention process. A radiology report is first prepared by radiologists for every patient study as part of the clinical routine. In the second stage, human annotators extract and confirm structured labels from these reports using artificial rules and existing natural language processing (NLP) tools.
However, there are two major limitations of this label extraction workflow. First, it is still complex and labor intensive. For example, human annotators have to define a list of alternate spellings, synonyms, and abbreviations for every target label. Consequently, the final accuracy of extracted labels heavily depends on the quality of human assistance and various NLP tools. A small mistake in a single step or a single tool may give rise to disastrous annotation results. Second, those human-defined rules are often severely restricted to application-oriented tasks instead of general-purpose tasks. It is difficult for DCNNs to learn universal representations from such application-oriented tasks.

In this paper, we propose REviewing FreE-text Reports for Supervision (REFERS) to directly learn radiograph representations from accompanying free-text radiology reports. We believe abstract and complex logic reasoning sentences in radiology reports provide sufficient information for learning well-transferable visual features. As shown in Figure \ref{workflow}a, REFERS is realized using a set of transformers, where the most important part is a radiograph transformer serving as the backbone. The main reason why we choose the transformer as the backbone in REFERS is that it not only exhibits the advantages of DCNNs, but also has been shown to be more effective~\cite{dosovitskiy2020image} because of the self-attention mechanism~\cite{vaswani2017attention}. Moreover, we have found that, in comparison to features generated from DCNNs, features from transformers are more compatible with textual tasks.

Different from aforementioned representation learning methodologies, REFERS performs cross-supervised learning and does not need structured labels during the pre-training stage. Instead, supervision signals are defined by automatically cross-checking the two different data modalities, radiographs and free-text reports. Considering in daily clinical routine, there is typically a free-text report associated with every patient study, which usually involves more than one radiographs. To fully utilize the study-level information in each report, we design a view fusion module based on an attention mechanism to process all radiographs in a patient study simultaneously, and fuse the resulting multiple features. In this way, the learned representations are able to preserve both study-level and image-level information. In contrast, only image-level information is addressed in traditional representation learning paradigms \cite{wang2017chestx,chen2019self,zhou2021models,haghighi2021transferable,zhou2020comparing} that use a single image as input. On top of the view fusion module, we conduct two tasks, i.e., report generation and study-report representation consistency reinforcement, to extract study-level supervision signals from free-text reports. To carry out the first task, we apply a decoder, called report transformer, to the fused feature with the goal to reproduce the radiology report associated with the study. For the second task, we apply our radiograph transformer and an NLP transformer to a study-report pair. These transformers produce a pair of feature representations for the patient study and radiology report in the pair, respectively. The consistency between such a pair of feature representations within every study-report pair is reinforced via a contrastive loss function. Some previous works~\cite{shin2015interleaved,wang2018tienet} tried to learn joint text-image representations for single-domain medical image analysis tasks. Compared to them, REFERS focuses on learning well-transferable image features from study-level free-text reports on a large-scale source domain and fine-tuning them on one or more target domains. 

On four well-known X-ray datasets, REFERS outperforms self-supervised learning and transfer learning on natural source images in producing more transferable representations, often bringing impressive improvements (more than 5\%) under limited supervision from target domains. This capability can be extremely important in real-world applications as medical data is scarce and their annotations are usually hard to acquire. More surprisingly, we found that REFERS clearly surpasses those methods that employ a source domain with a large collection of medical images with structured labels. In terms of specific abnormalities and diseases, REFERS is quite effective under extremely limited supervision ($<$ 1k annotated radiographs during fine-tuning). For instance, REFERS brings about 9-percent improvements on pneumothorax. Meanwhile, over 7-percent improvements are achieved on two common lung diseases (atelectasis and emphysema).

\section{Results}
All self-supervised learning (SSL) and label-supervised pre-training (LSP) baselines as well as our REFERS are first pre-trained on a source domain of medical images (i.e., MIMIC-CXR-JPG~\cite{johnson2019mimic}). Then, pre-trained models are fine-tuned on each of four well-established datasets (target domains with labels), including NIH ChestX-ray~\cite{wang2017chestx}, VinBigData Chest X-ray Abnormalities Detection~\cite{nguyen2020vindr}, Shenzhen Tuberculosis~\cite{jaeger2014two} and COVID-19 Image Data Collection~\cite{cohen2020covidProspective}. During the fine-tuning stage, we always perform fully-supervised learning on the target domain, which only consists of radiographs with structured labels. Furthermore, we verify model performance by varying the percentage of actually used training images (sampled from the predefined whole training set) in the target domain, and this percentage is called {\em label ratio}. When the label ratio is 100\%, we use the whole training set in the target domain for fine-tuning. \\

\begin{figure}
	\centering
	\includegraphics[width=0.80\columnwidth]{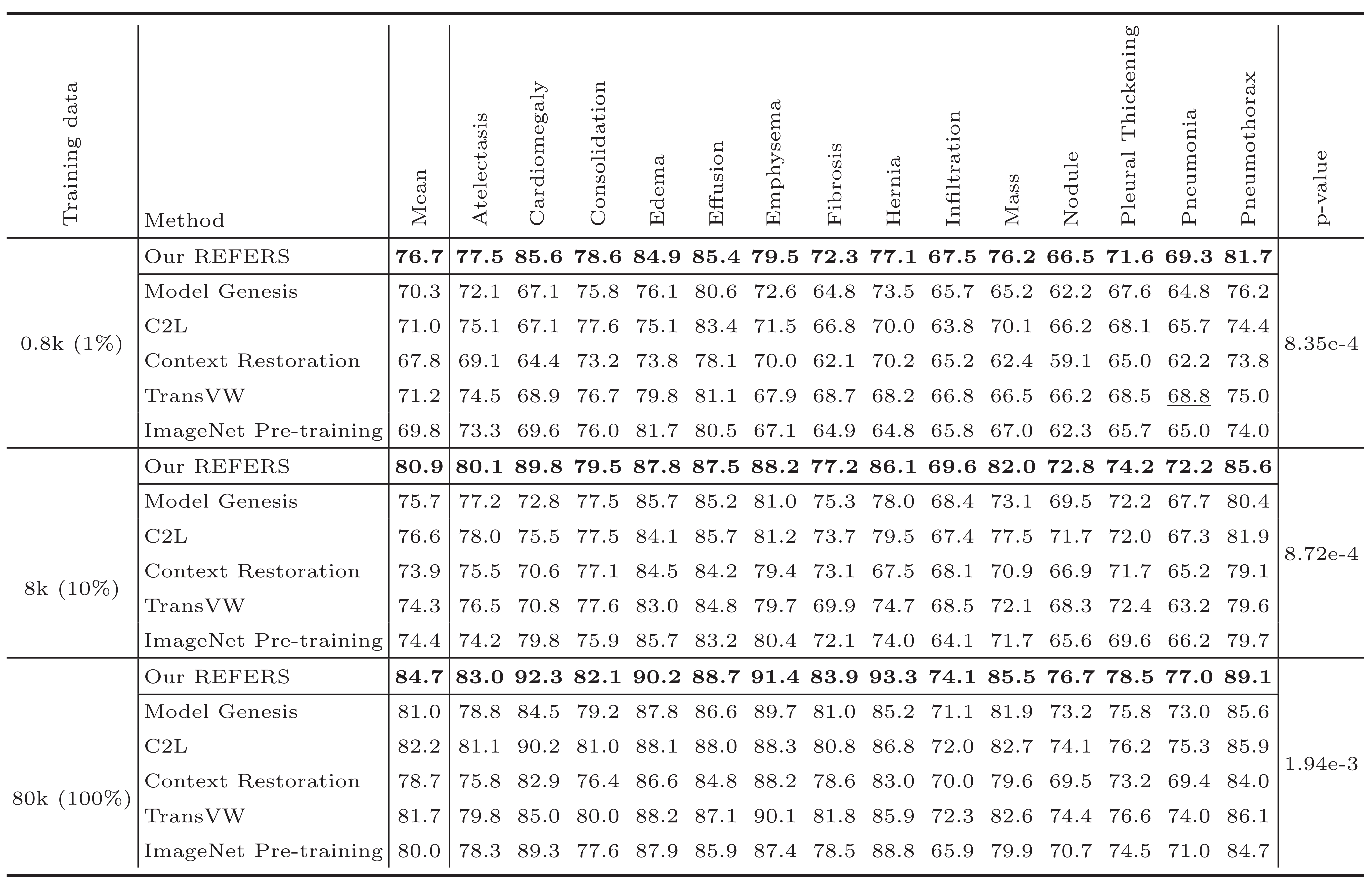}
	\caption{Comparison with self-supervised learning and transfer learning baselines on NIH ChestX-ray dataset. Note that for the sake of fairness, all baselines use the same transformer-based backbone as the radiograph transformer of REFERS (i.e., a ViT-like architecture plus the recurrent concatenation operator). Each p-value is calculated between our REFERS and the best performing baseline. The evaluation metric is Area under the ROC Curve (AUC). Best results are bolded.}
	\label{Ex_Fig_1}
\end{figure}

\begin{figure}
	\centering
	\includegraphics[width=0.80\columnwidth]{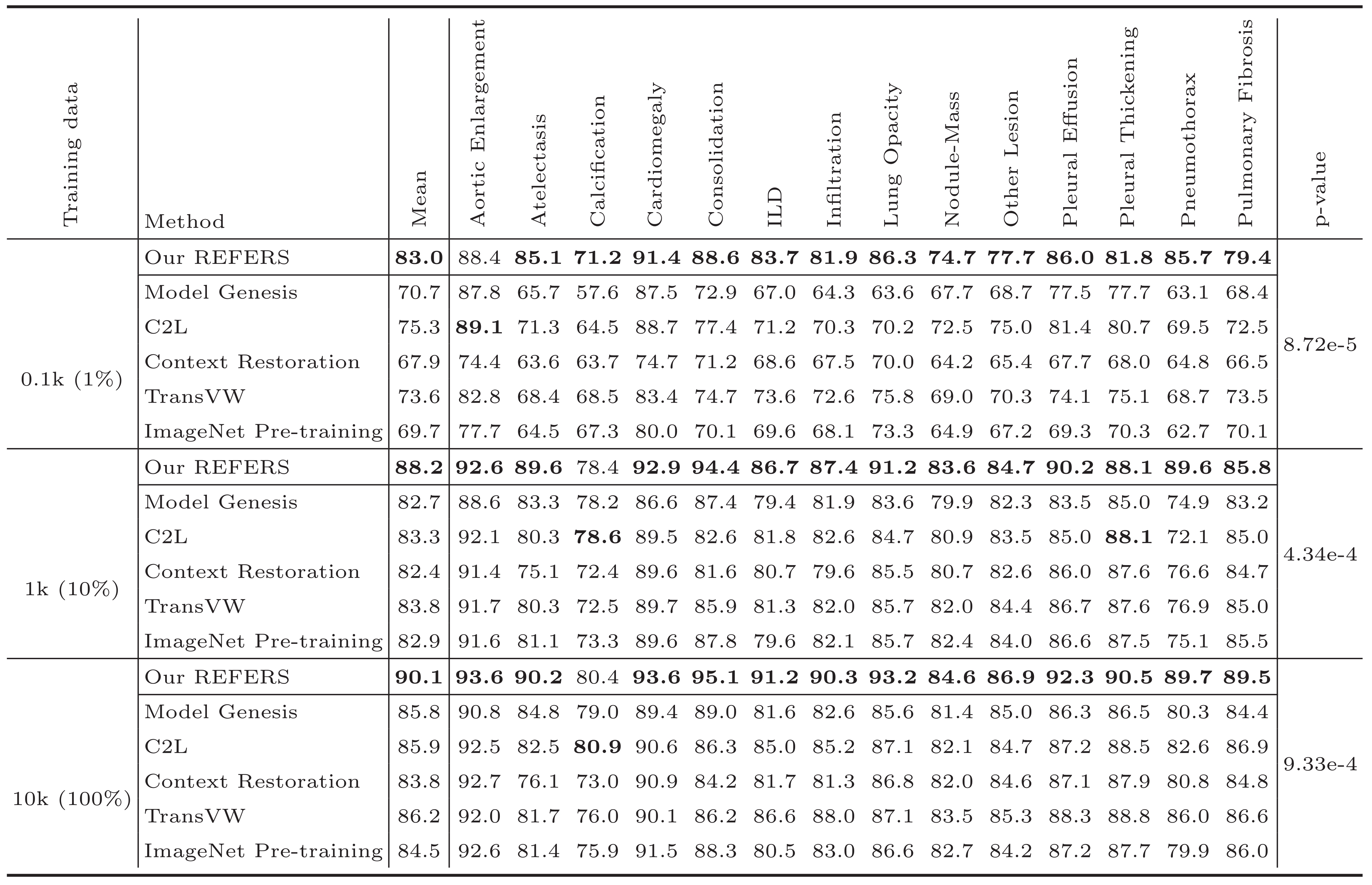}
	\caption{Comparison with self-supervised learning and transfer learning baselines on VinBigData Chest X-ray Abnormalities Detection. Note that for the sake of fairness, all baselines use the same transformer-based backbone as the radiograph transformer of REFERS (i.e., a ViT-like architecture plus the recurrent concatenation operator). Each p-value is calculated between our REFERS and the best performing baseline. The evaluation metric is Area under the ROC Curve (AUC). Best results are bolded.}
	\label{Ex_Fig_2}
\end{figure}

\noindent \textbf{NIH ChestX-ray.} Table~\ref{summary_base}, Extended Data Figures~\ref{Ex_Fig_1} and \ref{Ex_Fig_2} present experimental results from our REFERS and other approaches under different label ratios. As shown in Table~\ref{summary_base} and Extended Data Figure~\ref{Ex_Fig_1}, our approach significantly outperforms self-supervised baselines and transfer learning on natural source images. To be specific, REFERS achieves the highest AUC on all 14 classes using different amounts of training data during the fine-tuning stage. Moreover, REFERS shows the largest performance improvements with respect to these baselines when only 0.8k training images (1\% label ratio) in the target domain are utilized. For example, REFERS surpasses the widely adopted ImageNet-based pre-training~\cite{wang2017chestx} by about 7 percents on average. Even when compared to LSP, our REFERS still gives quite competitive results. In Table~\ref{summary_lsp}, it is easy to find out that the average performance of REFERS actually surpasses LSP, and consistently maintains an advantage of at least 2 percents. Compared to self-supervised baselines~\cite{chen2019self,zhou2021models,haghighi2021transferable,zhou2020comparing} and ImageNet-based pre-training~\cite{wang2017chestx}, REFERS achieves the largest improvements on emphysema (7 percents) and cardiomegaly ($>$ 10 percents), especially under limited supervision. When compared to LSP, our method achieves consistent improvements on mass ($>$ 4 percents).\\

\begin{figure}
	\centering
	\includegraphics[width=0.80\columnwidth]{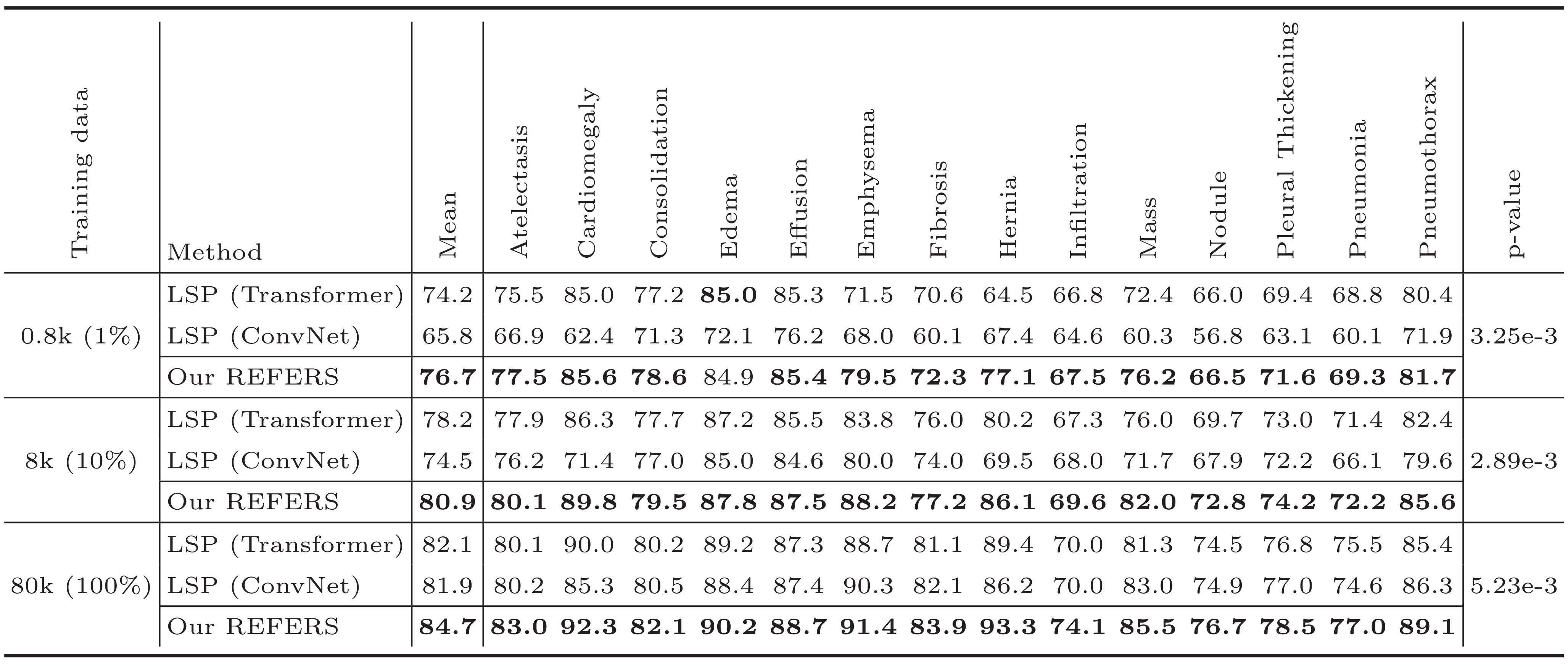}
	\caption{Comparison with label-supervised pre-training (LSP) on NIH ChestX-ray dataset. For fairness, both LSP (Transformer) and REFERS share the same transformer-based backbone (i.e., the ViT architecture plus the recurrent concatenation operator). Each p-value is calculated between the results from our REFERS and LSP (Transformer). The evaluation metric is Area under the ROC Curve (AUC). Best results are bolded.}
	\label{Ex_Fig_3}
\end{figure}

\begin{figure}
	\centering
	\includegraphics[width=0.80\columnwidth]{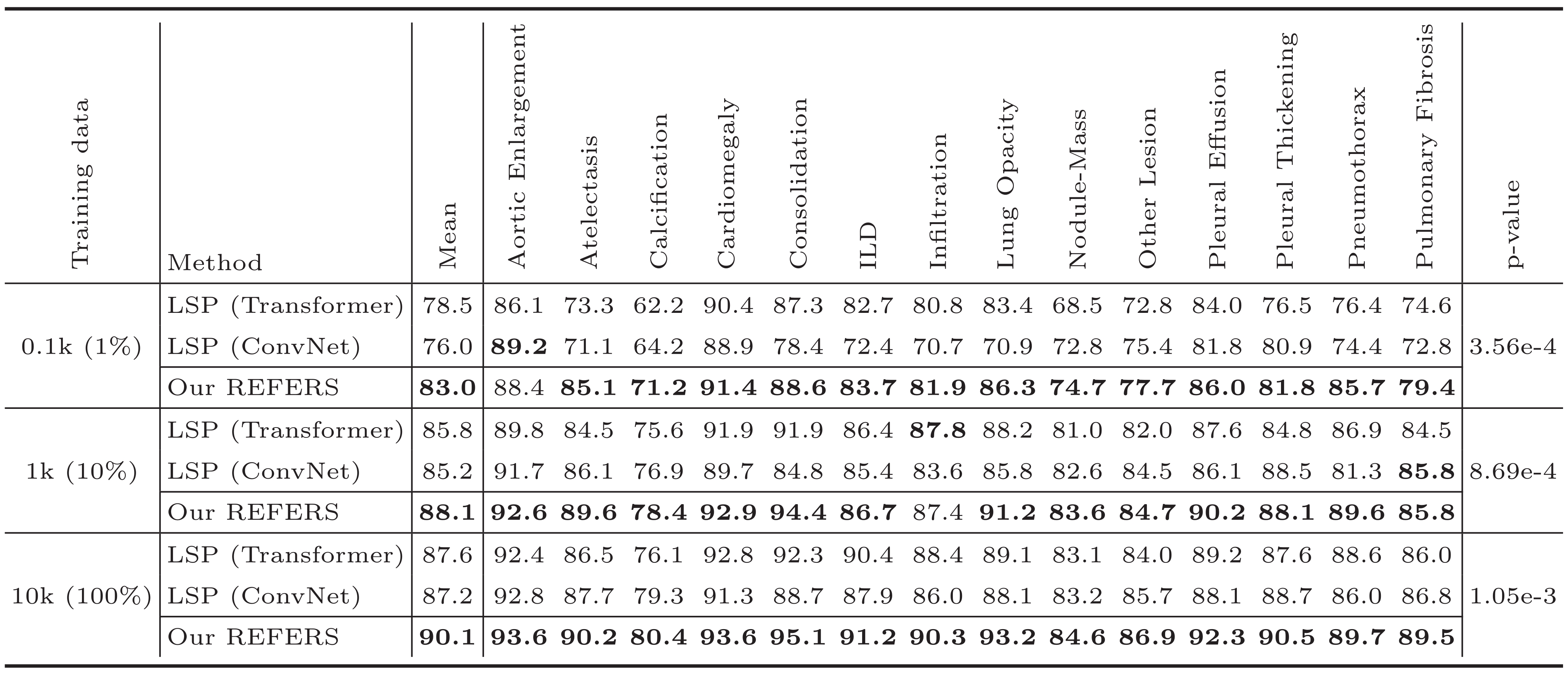}
	\caption{Comparison with label-supervised pre-training (LSP) on VinBigData Chest X-ray Abnormalities Detection. For fairness, both LSP (Transformer) and REFERS share the same transformer-based backbone (i.e., the ViT architecture plus the recurrent concatenation operator). Each p-value is calculated between the results from our REFERS and LSP (Transformer). The evaluation metric is Area under the ROC Curve (AUC). Best results are bolded.}
	\label{Ex_Fig_4}
\end{figure}

\noindent \textbf{VinBigData Chest X-ray Abnormalities Detection.} Our REFERS exhibits more advantage on this target domain dataset than it does on NIH ChestX-ray as VinBigData comprises a much smaller number of annotated radiographs (about $\frac{1}{8}$ of the NIH dataset). This phenomenon again demonstrates the ability of REFERS in dealing with limited supervision. REFERS consistently maintains large advantages over other methods under different conditions (see Tables~\ref{summary_base}, \ref{summary_lsp}, Extended Figures~\ref{Ex_Fig_3} and \ref{Ex_Fig_4}). For instance, when we only have 105 annotated radiographs (1\% label ratio) as fine-tuning data, REFERS surpasses C2L~\cite{zhou2020comparing}, the best performing self-supervised method, by over 7 percents in AUC. The performance of REFERS once again surpasses LSP with human-assisted structured labels even when all annotated training data (100\% label ratio) in the target domain is used. When we check specific abnormalities and diseases, we found REFERS consistently improves the diagnosis of atelectasis, lung opacity and pneumothorax in comparison to LSP.\\

\noindent \textbf{COVID-19 and Shenzhen Tuberculosis Image Collections} Both datasets serve as target domains and comprise a small number of labeled images (fewer than 1k X-rays), which are employed to test the transferability of the representation learned on the source domain. This is because few training images in such small target domains are not capable of training powerful models themselves. Thus, the performance of the trained models is more dependent on the quality of the learned representation. In Table~\ref{summary_base}, although separating tuberculosis from normal cases is not a hard task, our method still achieves 2.5\% improvements over C2L~\cite{zhou2020comparing} in AUC. When looking at COVID-19 Image Data Collection which includes two harder tasks, we can find that the relative performance improvements over self-supervised baselines~\cite{chen2019self,zhou2021models,haghighi2021transferable,zhou2020comparing} and transfer learning on natural source images~\cite{wang2017chestx} become quite clear. For instance, on the ``Viral vs. Bacterial” task, REFERS outperforms C2L~\cite{zhou2020comparing} by 7 percents in AUC, demonstrating the effectiveness of REFERS in helping achieve better performance over small-scale target datasets. Even if we compare REFERS against LSP, the performance advantage is still maintained at more than 1 percent. \\

\section{Discussion}

\noindent \textbf{REFERS outperforms self-supervised learning and transfer learning on natural source images by substantial and significant margins.} This is the most prominent observation obtained from our experimental results, which holds on different datasets and with different amounts of annotated training data during fine-tuning. Among self-supervised baselines~\cite{chen2019self,zhou2021models,haghighi2021transferable,zhou2020comparing}, C2L~\cite{zhou2020comparing} and TransVW~\cite{haghighi2021transferable} are the two best performing methods. Our REFERS outperforms C2L and TransVW by at least 4 percents when very limited annotated training data (at most 10\% label ratio) from NIH ChestX-ray and VinBigData datasets is used. Somewhat interestingly, as the label ratio increases, ImageNet-based pre-training~\cite{wang2017chestx} gradually narrows its gap with self-supervised learning. Nonetheless, our REFERS still surpasses it by a large margin (4 percents at least). Similar results can also be observed on Shenzhen Tuberculosis and COVID Image Collection. Since our REFERS employs a cross-supervised learning manner, it does not require structured labels as conventional fully-supervised learning approaches. As radiographs and radiology reports are readily available medical data, we believe our approach is as practical as self-supervised learning methodologies in real-world scenarios. \\

\noindent \textbf{REFERS consistently surpasses label-supervised pre-training with human-assisted structured labels.} This is another clear observation obtained from our experimental results. Even though our approach does not use any structured labels in the source domain, over all four target domain datasets, our pre-trained model exhibits clear advantages. Specifically, REFERS outperforms the most competitive LSP method, LSP (Transformer), which is based on Transformer and human-assisted structured labels in the source domain. In particular, our method shows more advantages at small label ratios. For instance, when NIH ChestX-ray and VinBigData are used as target domain datasets, REFERS achieves about 2.5\% improvements when the number of training images is smaller than 10k. Similarly, on Shenzhen Tuberculosis and COVID-19 Data Collection, REFERS consistently surpasses LSP by significant margins. It is worth mentioning that when a classification problem is difficult to solve and has limited supervision, REFERS becomes more advantageous and achieves impressive improvements. For example, on the ``Viral vs.  Bacterial" task (Table~\ref{summary_lsp}), REFERS surpasses label-supervised pre-training methods based on two-stage human intervention by approximately 4 percents. These improvements demonstrate that raw radiology reports contain more useful information than human-assisted structured labels. In other words, the advantages exhibited by our approach on small-scale target domain training data can be attributed to the rich information carried by radiology reports in the source domain. Such information provides additional supervision to help learn transferable representations for radiographs while the supervision signals from structured labels have less information. We believe this is an important step towards directly using natural language descriptions as supervision signals for image representation learning. As an example, our REFERS can be used to learn natural image representations from text descriptions at corresponding websites.\\

\begin{figure}[t]
	\centering
    \includegraphics[width=0.7\columnwidth]{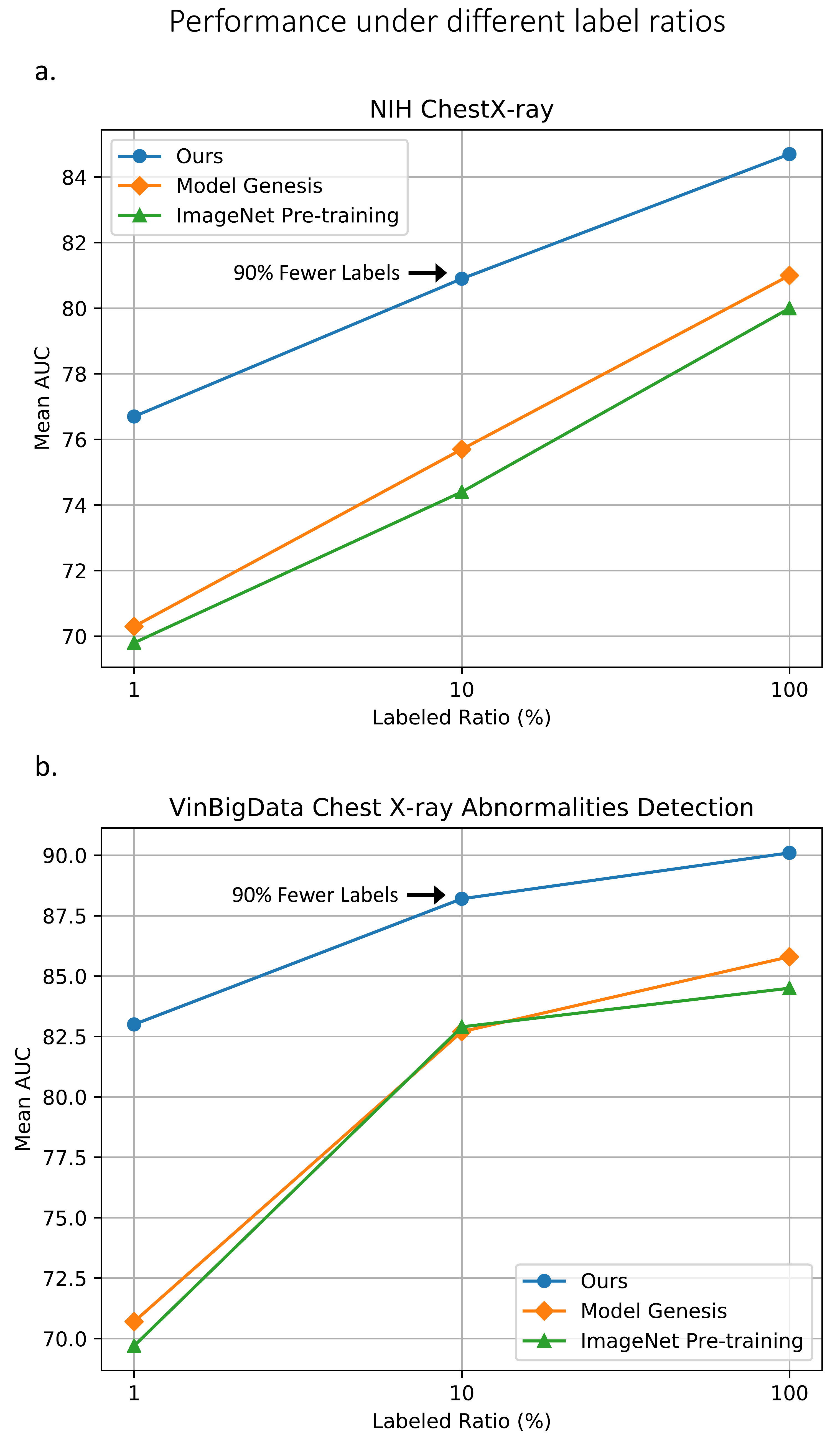}
    \caption{Performance obtained with different amounts of annotated training data in the target domain (a. NIH ChestX-ray and b. VinBigData Chest X-ray Abnormalities Detection). We also denote the percentage of annotated training data in the target domain that our REFERS requires to achieve comparable results with those of Model Genesis and ImageNet pre-training. Note that all three methods share the same transformer-based backbone.}
    \label{ratios}
\end{figure}

\noindent \textbf{REFERS significantly reduces the need of annotated data in target domains.} Figures~\ref{ratios}a and \ref{ratios}b present the performance of our approach under various label ratios. On NIH ChestX-ray, REFERS needs 90\% fewer annotated target domain data (10\% label ratio) to deliver a performance comparable to those of Model Genesis~\cite{zhou2021models} and ImageNet-based pre-training~\cite{wang2017chestx}. Similarly, on VinBigData, our method only needs 10\% annotated training data to achieve much better results than those of Model Genesis and ImageNet-based pre-training under 100\% label ratio. This phenomenon shows the potential of REFERS in providing high-quality pre-trained representations for downstream fine-tuning tasks with limited annotations. Due to the difficulty to acquire reliable annotations for medical image analysis, the ability to achieve good performance with limited annotations means much to the community.\\

\noindent \textbf{Improvements on specific abnormalities and diseases.} In Extended Data Figures~\ref{Ex_Fig_2}, REFERS brings 5-percent performance gains on emphysema and mass even when compared to LSP with limited supervision in the target domain ($<$ 10k training images). Since both abnormalities have a dispersed spatial distribution in the lung area, the considerable improvements demonstrate that REFERS is able to handle elusive chest abnormalities in radiographs well. When the amount of supervision in the target domain becomes extremely limited, such as using 105 training images from VinBigData, REFERS becomes more advantageous. For instance, REFERS outperforms LSP on atelectasis and pneumothorax by over 7 and 9 percents in Extended Data Figures~\ref{Ex_Fig_4}, respectively. Different from emphysema, mass and atelectasis, pneumothorax maintains a concentrated spatial distribution and is often located around the pleura. These successes imply that REFERS can deal with the diagnosis of both elusive and regular abnormalities and diseases well using a small number of training radiographs in the target domain. A similar phenomenon can be observed when REFERS is used for distinguishing viral pneumonia cases from bacterial ones in Tables~\ref{summary_base} and \ref{summary_lsp}. \\

\noindent \textbf{Transformer is more effective under limited supervision.} In Tables~\ref{summary_base} and \ref{summary_lsp}, we observe a trend of CNNs (i.e., ResNet series~\cite{he2016deep}): LSP (ConvNet) shows mediocre performance when a relatively small number of training images in the target domain are used. However, when all training data (100\% label ratio) is used, ConvNet shows competitive results. It seems that LSP (ConvNet) cannot well handle little amount of supervision. In contrast, LSP (Transformer) exhibits much better performance at small label ratios. This comparison demonstrates that pre-trained transformers generate more transferable representations than pre-trained CNNs. The underlying reason might be that the self-attention mechanism in transformers makes the learned representations more transferable due to captured long-distance dependencies.\\

\begin{figure}
	\centering
	\includegraphics[width=0.80\columnwidth]{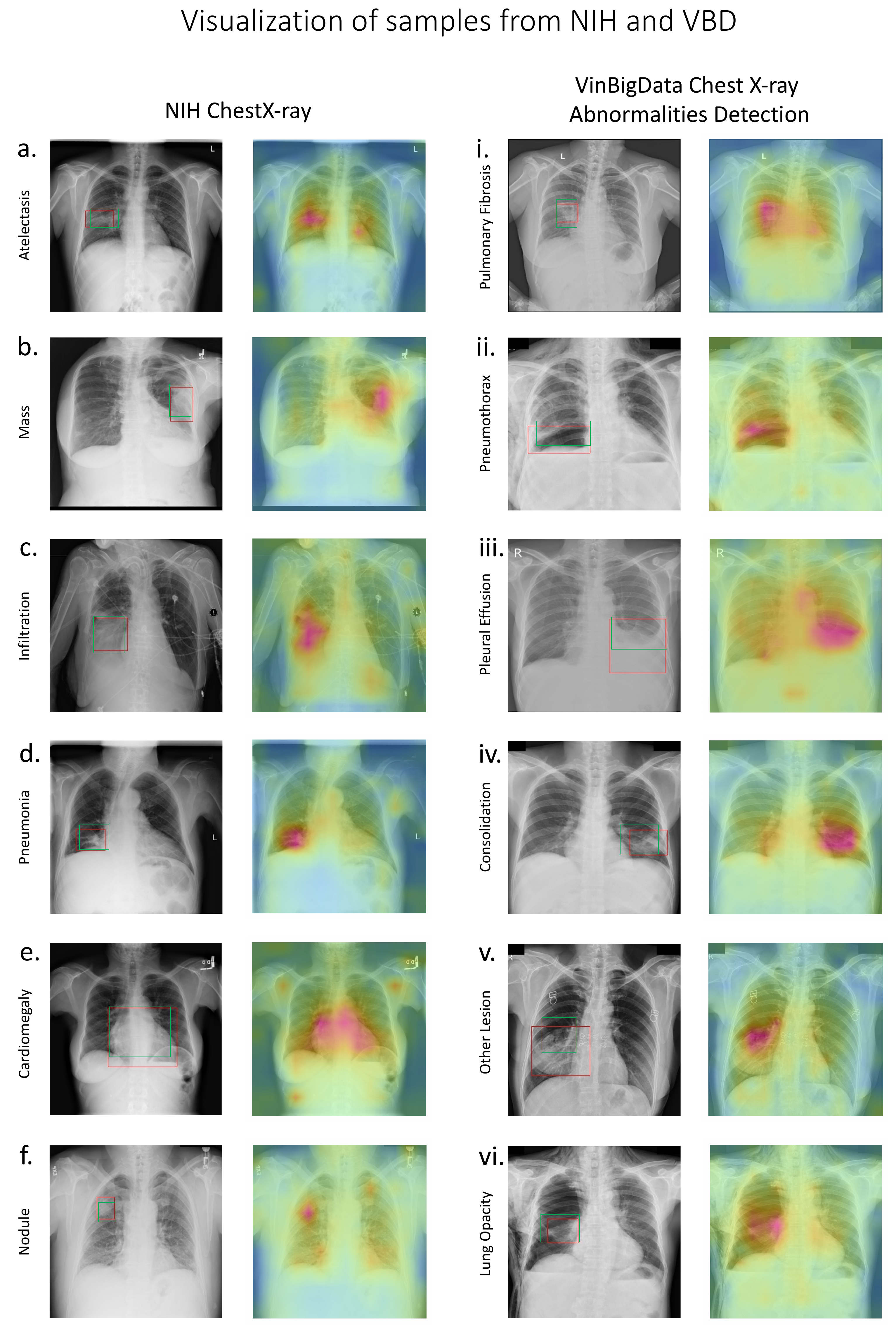}
	\caption{Visualization of twelve randomly chosen samples from NIH ChestX-ray (a-f) and VinBigData (i-vi) (fine-tuned with all annotated training data). For each sample, we present both the original image (left) and an attention map generated from REFERS. In each original image, red boxes denote lesion areas annotated by radiologists. In attention maps, fuchsia color stands for attention values generated from REFERS. The darker the fuchsia color, the higher the confidence of a specific disease. Green boxes in original images are our predicted lesion areas generated by applying a fixed confidence threshold to attention maps.}
	\label{vis_all}
\end{figure}

\noindent \textbf{REFERS provides reliable evidences for clinical decisions.} Figure~\ref{vis_all} presents randomly chosen radiographs and their corresponding class activation maps (CAMs)~\cite{zhou2016learning}. We can find that REFERS generates reliable attention regions, on top of which we can apply a fixed confidence threshold to further identify the location of different types of lesions (green boxes in Figure~\ref{vis_all}). The overall IoUs (Intersection over Unions) between green and red boxes (drawn by radiologists) are mostly higher than 0.5, indicating that the generated attention regions can well match radiologists' diagnoses. When lesions have a large size (such as the fifth image from NIH ChestX-ray), our method captures well-aligned lesion areas. Even when lesions are quite small and thus hard to detect (such as the last image from NIH ChestX-ray and the first image from VinBigData), REFERS can still identify the right locations. \\

\noindent \textbf{Replication of experimental results and their statistical significance.} There are a number of factors that influence pre-training results exhibit a certain level of randomness. These factors include, but are not limited to network initialization, training strategy (e.g., how to randomly crop images and perform mini-batch gradient descent) and even non-deterministic characteristics in computational tools (e.g., cuDNN~\cite{chetlur2014cudnn} would choose different algorithms in different runs due to benchmarking noise and hardware configuration). A good pre-training methodology should be able to produce relatively stable pre-trained representations when randomness in these factors is controlled within an acceptable limit. To take into account the influence of such randomness on experimental results, when REFERS and baseline pre-trained models are fine-tuned, we independently repeat each experiment three times and report their average results in Tables~\ref{summary_base} and \ref{summary_lsp}. Then, we calculate p-values between mean class AUCs of our REFERS and the best performing baseline model according to their fine-tuned performance using independent two-sample t-test. According to Tables~\ref{summary_base} and \ref{summary_lsp}, nearly all p-values are much smaller than 0.01, indicating that our REFERS is significantly better than its counterparts when various amounts of labeled training data in the target domain is used. In contrast, making the number of times (repeating each experiment) smaller than three would give rise to less stable mean AUCs while simply repeating more times would produce meaninglessly smaller p-values.

Last but not the least, we provide a thorough ablation study of REFERS in Table~\ref{ab_study}. More details can be found in the Methods section.

\section{Methods}
\noindent \textbf{Dataset for pre-training (source domain).} MIMIC-CXR-JPG \cite{johnson2019mimic} contains over 370k radiographs organized into patient studies, each of which may have one or more radiographs taken from different views or at different times for the same patient. Each patient study has one free-text radiology report, and each radiograph is associated with a set of abnormality/disease labels obtained from two-stage human-assisted intervention as mentioned above. There are two major sections in each report: Findings and Impressions. The Findings section includes detailed descriptions of important aspects in the radiographs while the Impressions section summarizes most immediately relevant findings. 

To acquire human-assisted structured labels for radiographs (i.e., two-stage human intervention), annotators need to first define a list of labels for abnormalities and diseases, including alternate spellings, synonyms, and abbreviations. On the basis of local contexts and existing NLP tools, mentions of labels in reports are classified as positive, uncertain, or negative. An aggregation procedure is further applied to aggregate multiple mentions of a single label. Uncertain labels need to be double-checked by radiologists.
 
As radiology reports were originally prepared by radiologists as part of the daily clinical routine, they can be regarded as freely available information that does not require extra human efforts in contrast to structured labels. In practice, we only keep the Findings and Impressions sections in the reports. Also, we remove all study-report pairs, where the text section has less than 3 tokens (words and phrases), from the dataset. This screening procedure produces 217k patient studies.\\

\noindent \textbf{Datasets for fine-tuning (target domains).} We do not require these datasets adopted for fine-tuning to have radiology reports. Instead, only human-assisted annotations are used during the fine-tuning stage. We follow the official split of NIH ChestX-ray, where the percentages of training, validation and testing sets are 70\%, 10\% and 20\%, respectively. The same set of ratios are also employed for VinBigData Chest X-ray, Shenzhen Tuberculosis and COVID-19 Image Data Collection to build randomly split training, validation and testing sets.
\begin{itemize}
	\item \emph{NIH ChestX-ray} is a dataset for multi-label classification of 14 chest abnormalities (i.e., Atelectasis, Cardiomegaly, Consolidation, Edema, Effusion, Emphysema, Fibrosis, Hernia, Infiltration, Mass, Nodule, Pleural Thickening, Pneumonia and Pneumothorax). There are over 100k frontal-view X-ray images of about 32k patients in NIH ChestX-ray, where labels of radiographs were extracted from associated reports following a similar procedure as that for MIMIC-CXR-JPG.
	\item \emph{VinBigData Chest X-ray} provides labels of 14 chest diseases (i.e., Aortic enlargement, Atelectasis, Pneumothorax, Lung Opacity, Pleural thickening, ILD, Pulmonary fibrosis, Calcification, Pleural effusion, Consolidation, Cardiomegaly, Other lesion, Nodule-Mass and Infiltration), and consists of 15k postero-anterior chest X-ray images. Here we did not use the test set in Kaggle, which does not provide any annotations. All images were labeled by a panel of experienced radiologists.
	\item \emph{Shenzhen Tuberculosis} is a small dataset containing 662 frontal chest X-ray images primarily from hospital clinical routine. 336 abnormal X-rays show various manifestations of tuberculosis, and the remaining 326 images are normal. We simply perform binary classification on this dataset.
	\item \emph{COVID-19 Image Data Collection} is a dataset involving more than 900 pneumonia cases with chest X-rays, which was built to improve the identification of COVID-19. We conduct experiments on two tasks, which are a) distinguishing COVID-19 from the rest and b) separating viral pneumonia cases from bacterial ones.
\end{itemize}

\noindent \textbf{Baselines and label-supervised pre-training.} Since our method does not need structured labels required by traditional fully-supervised learning, we compare it against four recent self-supervised learning methods~\cite{chen2019self,zhou2021models,haghighi2021transferable,zhou2020comparing} and ImageNet-based pre-training~\cite{wang2017chestx}:
\begin{itemize}
    \item \emph{Context Restoration}~\cite{chen2019self} repeats the operation of swapping two randomly chosen small X-ray patches for a fixed number of times, and the neural network is asked to restore each altered image back to its original version.
    \item \emph{Model Genesis}~\cite{zhou2021models} applies multiple types of distortions to the input X-ray, including local shuffling, non-linear transformation, in- and out-painting. Similar to Context Restoration, Model Genesis asks the model to reconstruct the original image from the distorted one.
    \item \emph{TransVW}~\cite{haghighi2021transferable} contrasts local X-ray patches to exploit the semantics of anatomical patterns while restoring distorted image contents.
    \item \emph{C2L}~\cite{zhou2020comparing} proposes to construct homogeneous and heterogeneous data pairs by mixing both images and features on top of MoCo~\cite{he2020momentum}. C2L outperforms MoCo by observable margins on multiple X-ray benchmarks.
    \item \emph{ImageNet-based pre-training}~\cite{wang2017chestx} is taken as a representative method that sets a large-scale dataset of annotated natural images as the source domain.
\end{itemize}
Note that all above baselines are implemented using the same transformer-based network architecture as our REFERS (i.e, a ViT architecture plus the proposed recurrent concatenation module). Such an implementation arrangement is meant to rule out the influence of network architectures on final performance and maintain fairness in experimental comparisons.

Finally, our approach is compared against label-supervised pre-training (LSP) that directly sets a large collection of X-ray images with human-assisted structured labels as the source domain. For better comparison, we implement LSP on top of both CNN and Transformer based backbone networks. Specifically, LSP (Transformer) adopts the same Transformer based network architecture as REFERS and the aforementioned self-supervised and ImageNet-based pre-training baselines. LSP (ConvNet) stands for the best performing residual network among ResNet-18, ResNet-50 and ResNet-101~\cite{he2016deep}. \\

\noindent \textbf{Data augmentation and image resizing.} 
During the \emph{pre-training} stage, we resize each radiograph in the source domain to 256$\times$256 pixels, and then apply random cropping to produce 224$\times$224 images. Random horizontal flip, random rotation (-10 to 10 degrees) and random grayscale (brightness and contrast) are also applied to generate augmented images. When using random horizontal flip, we change the words `left’ and `right’ in the accompanying radiology report accordingly. During the \emph{fine-tuning} stage, we apply the same set of data augmentation strategies, which are random cropping, random rotation, random grayscale and random horizontal flip, to all four target domain datasets. As in the pre-training stage, we resize each radiograph in a target domain to 256$\times$256, and then generate 224$\times$224 cropped and augmented radiographs as input images. \\

\noindent \textbf{Algorithm Overview.} REFERS performs cross-supervised learning on top of a transformer based backbone, called radiograph transformer. Given a patient study, we first forward its views to the radiograph transformer for extracting view-dependent feature representations. Next, we perform cross-supervised learning that acquires study-level supervision signals from free-text radiology reports. To this aim, it is necessary and essential to use view fusion to obtain a unified visual representation for an entire patient study because each radiology report is associated with a patient study but not individual radiographs within the patient study. Such fused representations are then used in two tasks during the pre-training stage: report generation and study-report representation consistency reinforcement. The first task takes the free texts in original radiology reports to supervise the training process of the radiograph transformer. The second task reinforces the consistency between the visual representations of patient studies and the textual representations of their corresponding reports.

\subsection{Radiograph Transformer} 
The radiograph transformer accepts image patches as inputs. We divide each image into a grid of 14$\times$14 cells, each of which has 16$\times$16 pixels. We then flatten each image patch to form a 1D vector of pixels, and feed it to the transformer. At the beginning of the transformer, a patch embedding layer linearly transforms each 1D pixel vector into a feature vector. This vector is concatenated with a position feature produced from a learnable position embedding to help clarify the relative location of each patch in the whole input patch sequence. The concatenated feature is then passed through another linear transformation layer to make its dimensionality the same as that of the final radiograph feature. At the core part of the radiograph transformer, we stack twelve self-attention blocks, which have the same architecture but independent parameters (Figure \ref{workflow}b). We first follow the practice in \cite{vaswani2017attention} to build a single self-attention block and then repeat its operations multiple times. In each block, we apply layer normalization~\cite{ba2016layer} before the multi-head attention and perceptron layers, after which residual connections are added to stabilize the training process. In the perceptron layer, we employ a two-layer perceptron with the Rectified Linear Unit (ReLU)~\cite{dahl2013improving} as the activation function. Moreover, we add an aggregation embedding, which is responsible for gathering the information from different input features. As shown in Figure~\ref{workflow}b, in the last layer, recurrent concatenation is performed to repeatedly concatenates the learned aggregation embedding with the learned representation of every patch. This is different from the operation in vision transformer (ViT)~\cite{dosovitskiy2020image}, which only concatenates the aggregation embedding with patch features once.

\subsection{Cross-supervised Learning}
There are two major components in cross-supervised learning: the view fusion module for producing study-level representations and two report-related tasks exploiting study-level information from associated free-text reports.

\begin{figure}[htp]
    \centering
    \includegraphics[width=1.0\columnwidth]{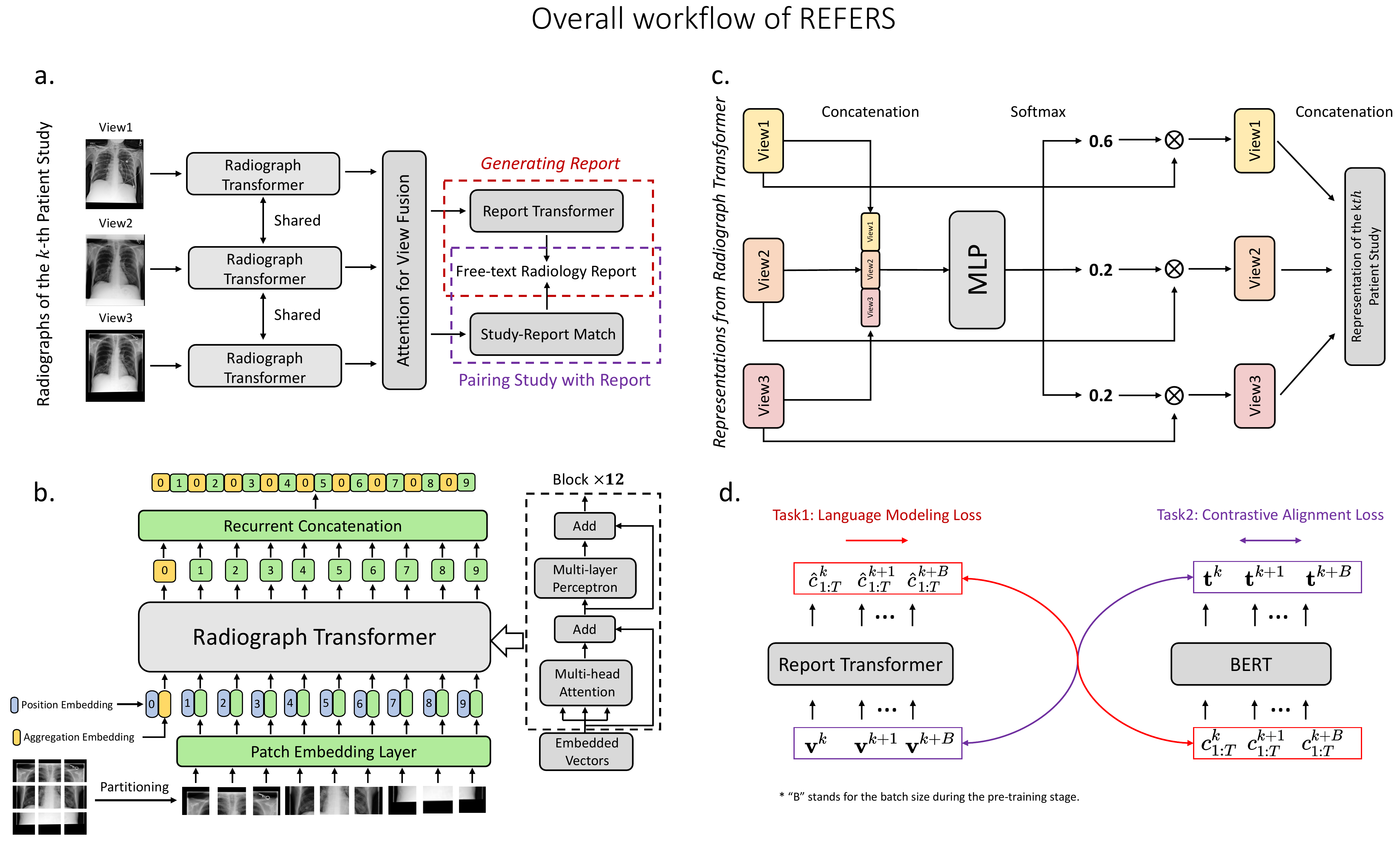}
    \caption{Workflow of REFERS: forwarding radiographs of the $k$-th patient study through the radiograph transformer, fusing representations of different views using an attention mechanism, and utilizing report generation and study-report representation consistency reinforcement to exploit the information in radiology reports. Part a provides an overview of the whole pipeline. Part b shows the architecture of the radiograph transformer. Attention for view fusion is elaborated in Part c. Part d presents two supervision tasks, report generation and study-report representation consistency reinforcement. In Part d, $\mathbf{v}^k$ and $\mathbf{t}^k$ denote the visual and textual features of the $k$-th patient study, respectively. $\hat{c}_{1:T}^k$ and $c_{1:T}^k$ stand for the token-level prediction and ground truth of the $k$-th radiology report whose length is $T$.}
    \label{workflow}
\end{figure}

As aforementioned, we forward all radiographs in a patient study through the radiograph transformer simultaneously to obtain their individual representations. We further employ an attention mechanism to fuse these individual representations to obtain an overall representation of the given study. Supposing a study has three radiographs (i.e., views), as shown in Figure~\ref{workflow}c. We first concatenate the features of all views, and then feed the concatenated features to a multi-layer perceptron to compute an attention value for each view. Next, we apply the softmax function to normalize these attention values, which are used as weights to produce a weighted version of the individual representations. Finally, these weighted representations are concatenated to form a unified visual feature for describing the whole study. Note that for studies that contain few than three radiographs, we randomly select one of the radiographs, and then repeat it once or twice to have a total of three views. For studies that contain more than three radiographs, we randomly select three of them from each study as input views.

We design two report-related tasks that acquire cross-supervision signals from free-text reports: report generation and study-report representation consistency reinforcement. In practice, these two tasks exploit study-level free-text information for better training study-level visual representations produced from the view fusion module. The first task applies a decoder, called report transformer, to the unified visual feature $\mathbf{v}^k$ of the $k$-th patient study to reproduce its associated radiology report denoted as $c_{1:T}^k$. Here, $c_1^k$ represents the start-of-sequence token and $c_T^k$ the end-of-sequence token. As a result, the report transformer generates a sequence of token-level predictions, $\hat{c}_{1:T}^k$, for the $k$-th patient study. The prediction of the $t$-th token in this sequence depends on the predicted subsequence $\hat{c}^k_{1:t-1}$ and the visual feature $\mathbf{v}^k$. The network architecture of the report transformer follows the architecture of the decoder in \cite{vaswani2017attention}. We wish the predicted token sequence ($\hat{c}_{1:T}^k$) resembles the sequence ($c_{1:T}^k$) representing the original report of the $k$-th patient study. Therefore, as shown in Figure~\ref{workflow}d, we apply a language modeling loss to both $\hat{c}_{1:T}^k$ and $c_{1:T}^k$ to maximize the following log-likelihood of the tokens in the original report.
\begin{equation}
\mathcal{L}^k_{\text{language}}=\sum_{t=2}^{T} \log P\left(c_{t}^k \mid \hat{c}_{1:t-1}^k, \mathbf{v}^k ; \phi_{v}, \phi_{t}\right),
\end{equation}
where $\hat{c}_{1}^k$ is a special symbol indicating the start of the predicted sequence, $\phi_v$ and $\phi_{t}$ stand for the parameters of the radiograph transformer and report transformer, respectively.

For the second task on study-report representation consistency reinforcement, we employ a contrastive loss~\cite{gutmann2010noise} to align cross-modal representations. Here, we use $\mathbf{t}^k$ to stand for the textual feature vector of the k-$th$ radiology report. In practice, we obtain $\mathbf{t}_k$ by forwarding the sequence of tokens in the k-$th$ report (i.e., $c^k_{1:T}$) to a BERT (i.e., Bidirectional Encoder Representations from Transformer) model~\cite{devlin2018bert}. BERT is built on top of the encoder in \cite{vaswani2017attention} using large-scale pre-training on a great number of corpus resources. Thus, BERT can help produce a generalized textual representation for the input report. Suppose we have $B$ patient studies in each training mini-batch, as shown in Figure~\ref{workflow}d. The contrastive loss for the k-$th$ study can be formulated as
\begin{equation}
	\mathcal{L}_{\text{contrast}}^k=-\log \frac{e^{\cos(\mathbf{v}^{k}, \mathbf{t}^{k}) / \tau}}{\sum_{i=1}^{B} e^{\cos(\mathbf{v}^{k}, \mathbf{t}^{i}) / \tau}},
\end{equation}
where $\cos(\cdot, \cdot)$ means the cosine similarity, $\cos(\mathbf{v}^{k}, \mathbf{t}^{k})=\frac{(\mathbf{v}^k)^{\top}\mathbf{t}^k}{\|\mathbf{v}^k\|\|\mathbf{t}^k\|}$, $\top$ denotes the transpose operation, $\|\cdot\|$ stands for L2 normalization, and $\tau$ is the temperature factor. Finally, for each patient study, we simply sum up $\mathcal{L}_{\text{contrast}}^k$ and $\mathcal{L}^k_{\text{language}}$ as the overall loss.
During the \emph{fine-tuning} stage, we typically use the cross entropy loss for model tuning. \\

\noindent \textbf{Training and testing methodologies.} We first pre-train the radiograph transformer on the source domain and then fine-tune it on downstream target domain datasets to verify the quality of pre-training. During the \emph{pre-training} stage, we sample 4.6k studies to form a held-out validation set according to the official division of the MIMIC-CXR-JPG dataset~\cite{johnson2019mimic}. We train the entire network using stochastic gradient descent (SGD) while setting the momentum value to 0.9~\cite{sutskever2013importance} and the weight decay to 1e-4. Following~\cite{devlin2018bert}, we do not apply weight decay to layer normalization and the bias terms in all layers. We use a fixed batch size of 32 for 300k iterations (about 45 epochs). We calculate the validation loss after each epoch and save the checkpoint that achieves the lowest validation loss. We adopt the linear learning rate warm-up strategy~\cite{goyal2019scaling} for the first 10k iterations, and then switch to cosine decay~\cite{loshchilov2016sgdr} until the end. Empirically, we found that training the radiograph transformer requires a large learning rate for fast convergence. Thus, its learning rate is set to 3e-3 while the learning rate for the report transformer and BERT is set to 3e-4. We initialize the aggregation embedding to all zeros while randomly initializing all position embeddings. We use PyTorch~\cite{paszke2019pytorch} and NVIDIA Apex for mixed{-}precision training~\cite{micikevicius2017mixed}. The complete pre-training process on the MIMIC{-}CXR dataset takes about 2 days on a single RTX 3090 GPU.

During the \emph{fine-tuning} stage, we fine-tune all transformer based models (including transformer based baselines) using SGD with the momentum set to 0.9 and the initial learning rate set to 3e-3 for all datasets. We fine-tune ResNet models using Adam~\cite{kingma2014adam} instead of SGD, and set the initial learning rate to 1e-4. All downstream models use the same learning rate decay strategy as that used in the pre-training stage, and are trained with a batch size of 128.

\subsection{Ablation Study}  
We conduct a thorough ablation study of REFERS by removing or replacing individual modules, and the results are shown in Table~\ref{ab_study}. 

First, we investigate the impact of replacing the radiograph transformer (rows 1-2 in Table~\ref{ab_study}). If we replace the radiograph transformer with ResNet-101~\cite{he2016deep} (row 1), the overall performance of REFERS on COVID-19 Image Data Collection would drop by about 7 percents (compared to row 0). This comparison demonstrates that the radiograph transformer is more effective in dealing with limited annotations, which is also verified with results in Tables~\ref{summary_base} and \ref{summary_lsp}. Next, when we replace the radiograph transformer with the original ViT architecture (row 2), which does not have the recurrent concatenation operator, the overall performance would drop by 3.3 percents. This result verifies the helpfulness of recurrently concatenating the learned aggregation embedding with patch representations. We also note that there exists a 3.8-percent performance difference between ResNet and ViT based architectures (rows 1\&2), showing the advantage of a transformer-like architecture. 

In addition to the radiograph transformer, we also investigate the impact of cross-supervised learning. First of all, we remove the view fusion module so that different radiographs within a patient study become associated with the same study-level radiology report (row 3). Such an operation is counter-intuitive as each individual radiograph alone cannot provide enough information to produce a study-level report. By comparing row 3 with row 0, we found that dropping the view fusion module would reduce the performance by nearly 2 percents on COVID-19 Image Data Collection. This result implies that learning study-level pre-trained representation is better than image-level pre-training as the former includes more patient-level information. Next, we completely replace cross-supervised learning with label-supervised learning (row 4), and REFERS deteriorates into LSP (Transformer) in Table~\ref{summary_lsp}. We found that dropping the two report-related tasks would adversely affect the performance by 2 percents. Last but not the least, we study the two report-related learning tasks individually. By comparing row 0 with row 5 and row 6, respectively, we observed that dropping either of them would not affect the overall performance too much (about 1 percent). This result implies that the effects of both tasks may partially overlap to some extent. Nonetheless, either of them along with the view fusion module can still outperform LSP (Transformer) (row 4). In addition, we found that although both of them improve the overall performance, reinforcing the consistency between representations of each patient study and its associated report (i.e., the second task) is more crucial than report generation (i.e., the first task). We believe the reason behind is that the representation learned in the second task can be regarded as a summary of each report, and thus provides more global information than token-level predictions in the first task. Such advantages make it more beneficial for the second task to include more study-level information for learning better study-level radiograph features.

\section*{Code Availability}
All codes are available at \url{https://github.com/funnyzhou/REFERS}~\cite{refers2021code}.

\section*{Data Availability}
\noindent \textbf{MIMIC-CXR-JPG}: \url{https://physionet.org/content/mimic-cxr-jpg/2.0.0/}.\\

\noindent \textbf{NIH ChestX-ray}: \url{https://nihcc.app.box.com/v/ChestXray-NIHCC/folder/36938765345}.\\

\noindent \textbf{VinBigData Chest X-ray Abnormalities Detection}: \url{https://www.kaggle.com/c/vinbigdata-chest-xray-abnormalities-detection}.\\

\noindent \textbf{Shenzhen Tuberculosis}: \url{https://www.kaggle.com/raddar/tuberculosis-chest-xrays-shenzhen}.\\

\noindent \textbf{COVID-19 Image Data Collection}: \url{https://github.com/ieee8023/covid-chestxray-dataset}.

\section*{Acknowledgements}
This work was supported in part by the Fundamental Research Funds for the Central Universities (Grant No. 20720190012, 20720210121).

\section*{Author Contributions Statement}

H.Z. and Y.Y. conceived the idea and designed the experiments. H.Z., X.C. and Y.Z. implemented and performed the experiments. H.Z. and Y.Y. wrote the manuscript. All authors analyzed the data and experimental results, commented on the manuscript. 

\section*{Competing Interests Statement}

The authors declare no competing interests.

\clearpage
\section*{Tables}

\vfill
\begin{table}[h]
    \centering
    \resizebox{1.0\textwidth}{!}{
    \begin{tabular}{l|c|c|c|c|c|c|c|c|c}
    \toprule
         & NIH & NIH & NIH & VBD & VBD & VBD & SZ & C-T1 & C-T2 \\ 
         \hline
         Method & 0.8k (1\%) & 8k (10\%) & 80k (100\%) & 0.1k (1\%) & 1k (10\%) & 10k (100\%) & All & All & All\\
         \hline
         Our REFERS & \textbf{76.7} & \textbf{80.9} & \textbf{84.7} & \textbf{83.0} & \textbf{88.2} & \textbf{90.1} & \textbf{98.0} & \textbf{82.1} & \textbf{80.4} \\
         \hline
         Model Genesis & 70.3 & 75.7 & 81.0 & 70.7 & 82.7 & 85.8 & 94.9 & 76.0 & 71.8 \\
         C2L & 71.0 & 76.6 & 82.2 & 75.3 & 83.3 & 85.9 & 95.5 & 77.8 & 73.0 \\
         Context Restoration & 67.8 & 73.9 & 78.7 & 67.9 & 82.4 & 83.8 & 92.7 & 74.6 & 69.8 \\
         TransVW & 71.2 & 74.3 & 81.7 & 73.6 & 83.8 & 86.2 & 94.2 & 76.1 & 71.5 \\
         ImageNet Pre-training & 69.8 & 74.4 & 80.0 & 69.7 & 82.9 & 84.5 & 94.5 & 74.1 & 70.3\\
         \hline
         p-value & 8.35e-4 & 8.72e-4 & 1.94e-3 & 8.72e-5 & 4.34e-4 & 9.33e-4 & 1.73e-3 & 5.88e-4 & 3.59e-4\\
    \bottomrule
    \end{tabular}}
    \caption{Comparison with self-supervised learning and transfer learning baselines. \textbf{NIH}, \textbf{VBD} and \textbf{SZ} stand for NIH ChestX-ray, VinBigData Chest X-ray Abnormalities Detection and Shenzhen Tuberculosis datasets, respectively. \textbf{C-T1} and \textbf{C-T2} denote the two tasks in COVID-19 Image Data Collection, where one task is to distinguish COVID-19 from the rest (C-T1) and the other task is to separate viral pneumonia cases from bacterial ones (C-T2). Note that for the sake of fairness, all baselines use the same transformer-based backbone as the radiograph transformer of REFERS (i.e., a ViT-like architecture plus the recurrent concatenation operator).
	Each p-value is calculated between our REFERS and the best performing baseline. The evaluation metric is Area under the ROC Curve (AUC). Best results are bolded.}
    \label{summary_base}
\end{table}
\vfill

\vfill
\begin{table}[h]
    \centering
    \resizebox{1.0\textwidth}{!}{
    \begin{tabular}{l|c|c|c|c|c|c|c|c|c}
    \toprule
         & NIH & NIH & NIH & VBD & VBD & VBD & SZ & C-T1 & C-T2 \\ 
         \hline
         Method & 0.8k (1\%) & 8k (10\%) & 80k (100\%) & 0.1k (1\%) & 1k (10\%) & 10k (100\%) & All & All & All\\
         \hline
         Our REFERS & \textbf{76.7} & \textbf{80.9} & \textbf{84.7} & \textbf{83.0} & \textbf{88.2} & \textbf{90.1} & \textbf{98.0} & \textbf{82.1} & \textbf{80.4} \\
         \hline
         LSP (Transformer) & 74.2 & 78.2 & 82.1 & 78.5 & 85.8 & 87.6 & 96.4 & 80.2 & 76.6 \\
         LSP (ConvNet) & 65.8 & 74.5 & 81.9 & 76.0 & 85.2 & 87.2 & 96.7 & 80.1 & 76.2 \\
         \hline
         p-value & 3.25e-3 & 2.89e-3 & 5.23e-3 & 3.56e-4 & 8.69e-4 & 1.05e-3 & 9.65e-3 & 7.61e-3 & 1.47e-3\\
    \bottomrule
    \end{tabular}}
    \caption{Comparison with methods using human-assisted structured labels. \textbf{NIH}, \textbf{VBD} and \textbf{SZ} stand for NIH ChestX-ray, VinBigData Chest X-ray Abnormalities Detection and Shenzhen Tuberculosis datasets, respectively. \textbf{C-T1} and \textbf{C-T2} denote the two tasks in COVID-19 Image Data Collection, where one task is to distinguish COVID-19 from the rest (C-T1) and the other task is to separate viral pneumonia cases from bacterial ones (C-T2). Note that for fairness, both LSP (Transformer) and REFERS share the same transformer-based backbone (i.e., the ViT architecture plus the recurrent concatenation operator). Each p-value is calculated between the results from our REFERS and LSP (Transformer). The evaluation metric is Area under the ROC Curve (AUC). Best results are bolded.}
    \label{summary_lsp}
\end{table}
\vfill

\clearpage
\vfill
\begin{table}[]
    \centering
    \resizebox{1.0\textwidth}{!}{
    \begin{tabular}{c|c|c|c|c|c|c}
    \toprule
    Row & ViT & RecConcate & View Fusion & Task1 & Task2 & Viral vs. Bacterial\\
    \hline
    0& \checkmark & \checkmark & \checkmark & \checkmark & \checkmark & \textbf{80.4} \\
    \hline
    1& & & \checkmark & \checkmark & \checkmark & 73.3  \\
    2& \checkmark &  & \checkmark & \checkmark & \checkmark & 77.1 \\
    \hline
    3& \checkmark & \checkmark & & \checkmark & \checkmark & 78.6\\
    4& \checkmark & \checkmark & & & & 76.6\\
    5& \checkmark & \checkmark & \checkmark & \checkmark & & 79.1\\
    6& \checkmark & \checkmark & \checkmark & & \checkmark & 79.3\\
    \bottomrule
    \end{tabular}}
    \caption{An ablation study of REFERS by removing or replacing individual modules. \textbf{RecConcate} stands for the recurrent concatenation operation in the radiograph transformer. \textbf{Task1} and \textbf{Task2} refer to the two tasks in cross-supervised learning. Row 1 corresponds to the result of a convolutional neural network while row 4 corresponds to LSP (Transformer).}
    \label{ab_study}
\end{table}
\vfill

\clearpage
\section*{Figures}

\vfill
\vfill

\vfill
\vfill

\vfill

\vfill
\vfill

\vfill
\vfill

\vfill
\vfill

\vfill

\end{document}